\documentclass[prl,twocolumn,floatfix,amsfonts]{revtex4}
\usepackage{graphicx,graphics,color,epsfig}% Include figure files
\usepackage{bm}
\usepackage{amsmath}
\usepackage{amssymb}
\usepackage{color}

\begin{document}
\title{Comment on
``Zeeman-Driven Lifshitz Transition: A Model for the
Experimentally Observed Fermi-Surface Reconstruction in ${\bf {\rm
YbRh_2Si_2}}$"}
\author{S. Friedemann$^{1}$,
S. Paschen$^2$,
C. Geibel$^1$,
S. Wirth$^{1}$,
F. Steglich$^1$,
S. Kirchner$^{3}$,
E. Abrahams$^4$,
Q. Si$^{5}$ \\
}

\affiliation{$^{1}$Max Planck Institute for Chemical Physics of
Solids, N{\"o}thnitzer Str.~40, D - 01187~Dresden, Germany\\
$^{2}$Institute of Solid State Physics, TU Vienna, Wiedner
Hauptstra{\ss}e 8-10, 1040 Vienna, Austria \\
$^3$ Max Planck Institute for the Physics of Complex Systems,
N{\"o}thnitzer Str.~38, 01187~Dresden, Germany\\
$^4$Department of Physics and Astronomy, UCLA, Los Angeles, CA 90095, USA\\
$^{5}$Department of Physics and Astronomy, Rice University, Houston,
Texas 77005, USA
}
\begin{abstract}
In Phys. Rev. Lett. 106, 137002 (2011), A. Hackl and M. Vojta have proposed to explain
the quantum critical behavior of ${\rm YbRh_2Si_2}$ in terms of a Zeeman-induced
Lifshitz transition of an electronic band whose width is about 6 orders
of magnitude smaller than that of conventional metals. Here, we note that
the ultra-narrowness of the proposed band, as well as the proposed scenario {\it per se},
lead to properties which are qualitatively inconsistent with the salient features
observed in ${\rm YbRh_2Si_2}$ near its quantum critical point.
\end{abstract}
\maketitle

The field-induced transition out of an antiferromagnetic (AF)
order in ${\rm YbRh_2Si_2}$ has emerged as a prominent example of
local Kondo-destruction quantum critical point (QCP)
\cite{Friedemann-pnas,Paschen-nature,Gegenwart-science}. This
class of QCP goes beyond the Landau framework of order-parameter
fluctuations, involving a localization-delocalization transition
of the f-electrons across the AF transition; the Fermi surface
experiences a sudden jump from ``small" (f-electrons being localized) to
``large" (f-electrons delocalized) at zero temperature
\cite{Si_etal,Coleman_etal}. In ${\rm YbRh_2Si_2}$, the
evolution of the Hall coefficient as a function of magnetic field
provides the most direct evidence for 
%a sudden jump of the Fermi
%surface at zero temperature 
such a Fermi-surface jump 
across a field-driven QCP
\cite{Friedemann-pnas,Paschen-nature},
% this evidence 
which is
corroborated by the evolutions of the magnetoresistivity
\cite{Friedemann-pnas,Paschen-nature} and thermodynamic
properties  \cite{Gegenwart-science}. 
%The jump at $T=0$ of the
%Fermi surface 
This jump in the Fermi surface 
across the AF critical field $B_N$,
% = B^*(T\!\rightarrow \! 0)$, 
which is about $0.06$ T for a field
applied within the $ab$ plane of the tetragonal crystalline
structure, appears to be accompanied by the divergence of the quasiparticle
mass \cite{Gegenwart_prl}.

In a recent paper \cite{HV_prl}, Hackl and Vojta (HV) proposed an
alternative interpretation of these observations in ${\rm
YbRh_2Si_2}$ near $B_N$, based on a Lifshitz transition associated
with the bandstructure of non-interacting electrons or weakly-interacting quasiparticles. 
HV postulated a hitherto unknown ultra-narrow band, with one
component of its Zeeman-split Fermi surface shifting below
(or above) the Fermi energy at the critical field. (In the
standard classification of Lifshitz transitions, this belongs to
the ``void" type \cite{Lifshitz_review}.) Because the critical field in this case is very
small (in contrast to other cases of heavy fermions where a
field-induced Lifshitz transitions have been implicated
\cite{Huxley_natphys,Julian_prl}), HV had to assume an exceedingly
small bandwidth of about 5 $\mu$eV.

Here, we argue that the scenario of HV is unlikely to be pertinent
to the physics near $B_N$ in ${\rm YbRh_2Si_2}$. A number of key
features in the HV proposal contradict the salient experimental results.

{\it A.}~~Clearly, for a bandwidth of 5 $\mu$ eV invoked in the scenario of HV,
temperatures
above about 50 mK remove the underlying
Fermi surface and would therefore wash out the features of the
Lifshitz transition in the isothermal Hall coefficient and
magnetoresistivity. This is in contradiction to the experiment, in
which the isothermal crossover can be traced up to about 20 times as high 
a temperature (about 1 K), as
illustrated by Fig.~1a (Ref.~\cite{Friedemann-pnas}).

{\it B.}~~The 5 $\mu$eV bandwidth is so small that it would lead to
an entropy crisis. 

{\it B1.} 
Considering that 5 $\mu$eV is  about 
6 orders of magnitude smaller than the typical bandwidth of a simple metal,
the scenario of HV would very generally yield a huge 
specific-heat coefficient, $\gamma \equiv C_{el}/T$.
For example, the calculation reported in Fig.~3 of Ref.~\cite{HV_prl} is based on 
a full conduction-electron band (which has an entropy of the order of 
R$\ln 2$) of width about 5 $\mu$eV. The corresponding $\gamma$
will be about $10^3$ ${\rm J/mol \cdot K^2}$, which is 6 orders
of magnitude larger than that of a simple metal. 
Compared to the value observed experimentally at around 50 mK
(about $1$ ${\rm J/mol \cdot K^2}$) \cite{Gegenwart_prl}, this is 3 orders of magnitude 
too large. Equivalently, while the experimental observation has established that the entropy
associated with the quantum critical regime of ${\rm YbRh_2Si_2}$,
about $0.4\,$R$\ln 2$, is distributed in a temperature range of about 10K,
the scenario of HV would spread a similarly large entropy over a temperature 
range of about 50 mK.
 
{\it B2.} 
In a separate calculation, reported in their
Fig.~2, HV took the small-width band as riding on top of a
background portion corresponding to quasiparticles with a
Kondo scale of about 25 K. 
Because the invoked narrow band contains about 25\% of the spectral weight 
of the full band, the involved entropy is still too large -- 25\% of R$\ln 2$
over a temperature range of 50 mK. 
The $\gamma$ expected at around 50 mK
will still be much (25\% of three orders of magnitude) too large compared with 
the experimental observation.
Thus, this does not resolve the above entropy crisis.

{\it B3.} These order-of-magnitude issues should not be taken as quantitative 
problems. Instead, they represent qualitative discrepancies with the experiments
that are inherent to the assumed ultra-low Fermi energy scale. The background
density-of-states (DOS)  is associated with the 25 K energy scale,
which already corresponds to a $\gamma$ of the order 1 ${\rm J/mol \cdot K^2}$.
The proposed small peak of the DOS in the HV scenario can then only be of a 
similar height as the background one; in other words, it would hardly be a peak. 
Equivalently, its spectral weight must be so small (with the associated entropy
on the order of 0.2\% R $\ln 2$) 
that it will hardly have any observable consequence.

{\it C.}~~We next turn to the specific critical behavior. 
A void Lifshitz transition, with the Fermi surface
continuously shrinking to zero,  will lead to only weak singularities in the resistivity
and Hall effect as a function of the control parameter. In two dimensions, it typically
yields a cusp in the Hall coefficient and a change
in slope in the residual resistivity. In three dimensions, the features are even
weaker. In the scenario of HV, therefore, the
isothermal Hall coefficient and
magnetoresistivity will both 
%be smooth 
remain continuous
across the Zeeman-induced Lifshitz transition
down to $T=0$.
 Experimentally, however, both 
 %the isothermal Hall coefficient and electrical resistivity 
 quantities in ${\rm YbRh_2Si_2}$ have been shown to jump in the extrapolated $T=0$ limit:
%as a function of the magnetic field show a jump in the $T=0$
%limit: both quantities 
they
exhibit a crossover from one constant to
another at $B^*(T)$, and the crossover width $\Gamma$ goes to zero in
the $T=0$ limit. It should be stressed that, while experimental measurements are 
of course done at finite temperatures, our extrapolation of $\Gamma$ to the $T=0$ limit
is based on the observation that it is proportional to $T$ for more than
1.5 decades of temperature as shown in Fig.~1a
\cite{Friedemann-pnas}.

\begin{figure}
\begin{center}
\includegraphics[width=0.8\linewidth]{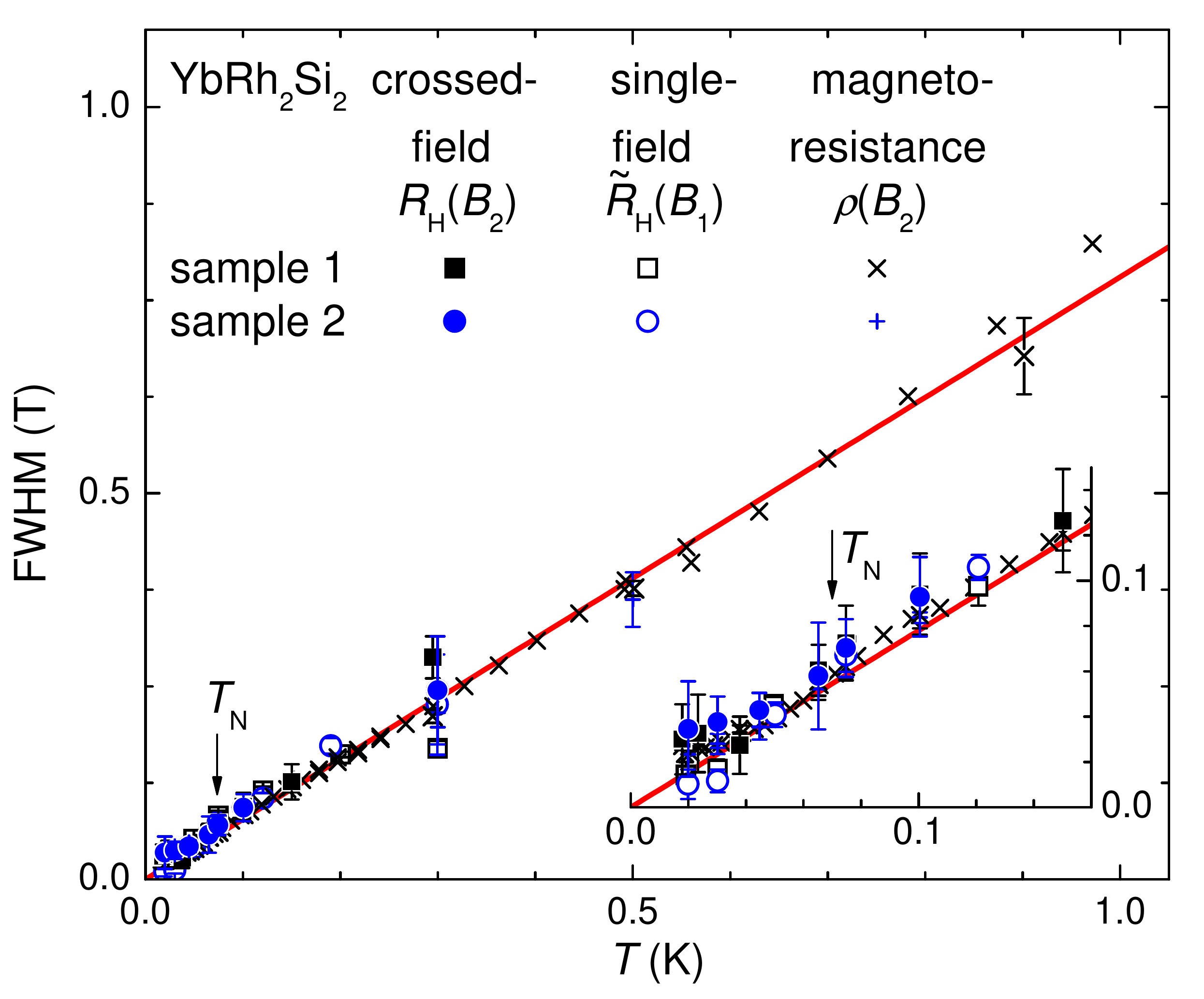}
\includegraphics[width=0.8\linewidth]{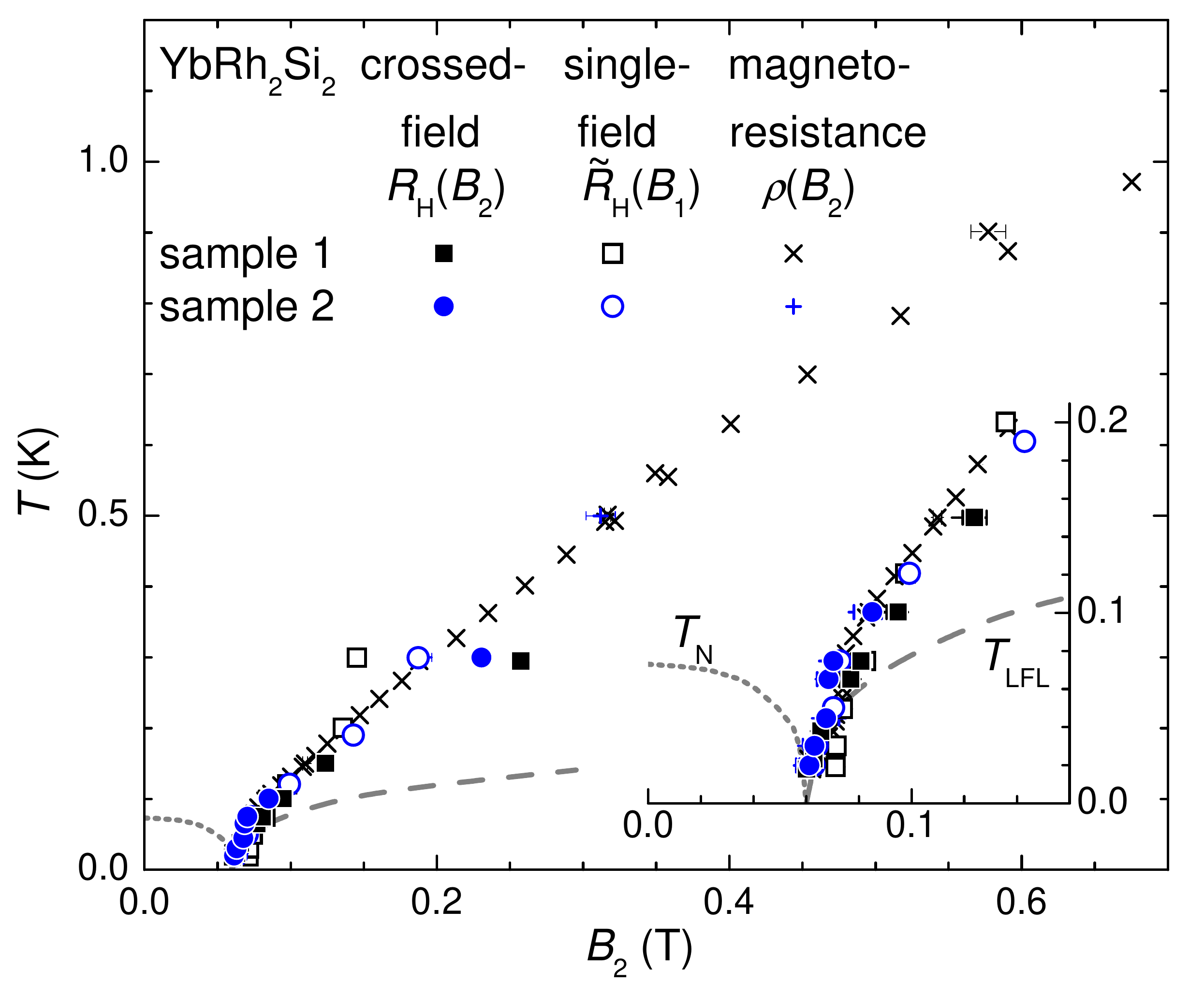}
\end{center}
\caption{%
(a) Full width at half maximum (FWHM, $\Gamma$) of the crossovers in the
Hall coefficient and magnetoresistivity. The solid line represents
a linear fit to all data sets. Within the error this line
intersects the origin. The inset magnifies data at the lowest
temperatures. The arrow indicates the N\'eel temperature at $B=0$; 
(b) Position of the crossover fields  in the temperature--magnetic
field phase diagram, $B^*(T)$, as determined from measurements
of the Hall coefficient and longitudinal magnetoresistivity. The inset
magnifies the low temperature range. The dotted and dashed lines
respectively represent the 
%boundary of the AF ordered state 
N\'eel temperature
and the crossover to low-$T$ Landau Fermi-liquid (LFL)
behavior on the paramagnetic side, as deduced from resistivity 
measurements.
(From Ref.~\cite{Friedemann-pnas}.) } \label{yrs}
\end{figure}

{\it D.}~~Because the void Lifshitz transition simply involves a
continuous shrinking of a Fermi pocket, 
it will produce neither a maximum nor a divergence of the effective 
mass at the transition point.
Indeed, in HV's calculations, $\gamma$ evolves monotonously and continuously through
the Lifshitz transition (Fig. 2b of Ref.~\cite{HV_prl}).
(In other heavy-fermion settings, experiments have implicated 
a Zeeman-induced Lifshitz transition 
at a high field \cite{Huxley_natphys,Julian_prl},
with the
effective mass {\em decreasing} on approach of the transition.)
This is in drastic contrast to the experiments in ${\rm YbRh_2Si_2}$:
both the enhancement and divergence  of the effective mass have been observed.
This divergence is most clearly inferred from the resistivity $A-$coefficient on both sides
of the critical field \cite{Gegenwart_prl}. It has also been observed through the measurement
of the specific-heat coefficient $\gamma$ on the elevated-field side \cite{Gegenwart_prl}, 
and is consistent with the behavior of the measured $\gamma$ in the low-field regime \cite{Brando}.

We close by noting that the most striking feature of the
phase diagram of YbRh$_2$Si$_2$ is that the $B^*(T)$
line terminates at the AF phase boundary as $T \rightarrow
0$, {\it i.e.} the AF critical 
$B_N$ equals $B^*(T=0)$,
% = B_N$, 
as shown in Fig.~1b
(Ref.\ \cite{Friedemann-pnas}), and there is evidence that this
property extends to a range of finite (negative)
chemical pressure as seen in Fig.~2 (Ref.\ \cite{Friedemann_jpsj}). It
is hard to imagine how this could emerge out of a fine-tuned
bandstructure of 
non-interacting electrons  
with no connection to the AF transition.

\begin{figure}
\begin{center}
\includegraphics[width=1.0\linewidth]{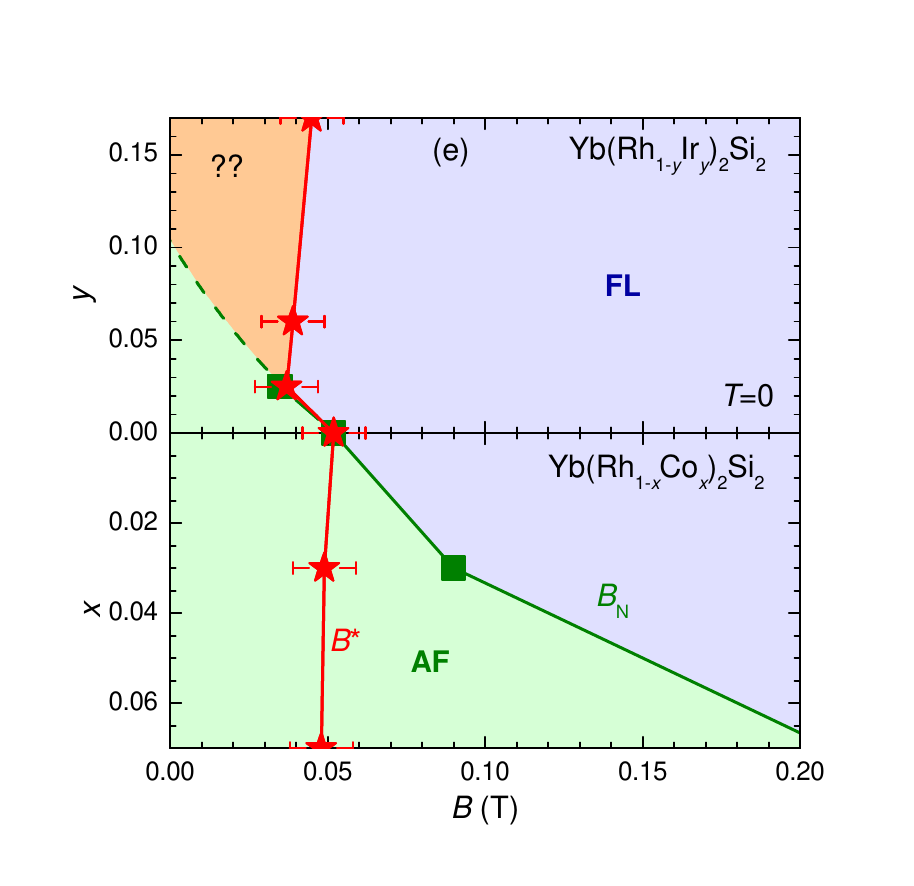}
\end{center}
\caption{%
Zero-temperature phase diagram of Yb(Rh$_{\rm 1-x}$M$_{\rm
x}$)$_{\rm 2}$Si$_{\rm 2}$ (M = Ir, Co). 
$B^*$ and $B_N$ are respectively the low-temperature limit of the crossover field
and the field for the AF transition;
their variations
% of $B^*$ and $B_N$ 
with composition are depicted. Green and blue shaded
region mark the AF ordered and paramagnetic Fermi-liquid
ground state, respectively. (From Ref.~\cite{Friedemann_jpsj}.)}
\label{doped}
\end{figure}

\end{document}